# Factors influencing graphene growth on metal surfaces

E. Loginova[1], N. C. Bartelt[1], P. J. Feibelman[2] and K. F. McCarty[1*]

Sandia National Laboratories
[1]Livermore, CA and [2]Albuquerque, NM

**Abstract**

Graphene forms from a relatively dense, tightly-bound C-adatom gas, when elemental C is deposited on or segregates to the Ru(0001) surface. Nonlinearity of the graphene growth rate with C adatom density suggests that growth proceeds by addition of C atom clusters to the graphene edge. The generality of this picture has now been studied by use of low-energy electron microscopy (LEEM) to observe graphene formation when Ru(0001) and Ir(111) surfaces are exposed to ethylene. The finding that graphene growth velocities and nucleation rates on Ru have precisely the same dependence on adatom concentration as for elemental C deposition implies that hydrocarbon decomposition only affects graphene growth through the rate of adatom formation; for ethylene, that rate decreases with increasing adatom concentration and graphene coverage. Initially, graphene growth on Ir(111) is like that on Ru: the growth velocity is the same nonlinear function of adatom concentration (albeit with much smaller equilibrium adatom concentrations, as we explain with DFT calculations of adatom formation energies). In the later stages of growth, graphene crystals that are rotated relative to the initial nuclei nucleate and grow. The rotated nuclei grow much faster. This difference suggests first, that the edge-orientation of the graphene sheets relative to the substrate plays an important role in the growth mechanism, and second, that attachment of the clusters to the graphene is the slowest step in cluster addition, rather than formation of clusters on the terraces.

## 1. Introduction

Formation of graphitic carbon has attracted attention for decades because of its importance in processes such as combustion, catalyst deactivation and degradation of multilayer mirrors in extreme ultraviolet lithography [1-3]. Interest in graphene, single-layer sheets of graphitic carbon, is rapidly growing owing to its potential use in next-generation electronics [4-8]. Graphene sheets have been prepared and characterized on many surfaces [9-20]. Graphene growth on metals is also widely studied to gain knowledge about the growth of carbon nanotubes from metal catalysts [21-23].

Preparation of epitaxial graphene on metal surfaces (e.g., Ru(0001) [12; 18], Ir(111) [10; 20; 15], Pt(111) [11] and Ni(111) [17]) typically employs three methods: segregation of C from the bulk of the metal to its surface, C-vapor deposition, and chemical-vapor deposition of hydrocarbons [11; 12; 18]. Wintterlin and Bocquet [24] have just reviewed graphene on metal surfaces, and concluded that graphene on some metals forms the most perfect supported films known. Still defects occur in the graphene films. For example, Coraux et al. have characterized small-angle misorientation boundaries of graphene on Ir(111) [9]. Graphene on Pt(111) forms several coexisting phases that differ in their in-plane orientation [11; 25]. The major focus in the literature has been the physical and electronic structure of the films. But, the nucleation and growth mechanisms, which can greatly influence the crystalline perfection of the films, remain largely unexplored experimentally.

Previously we used low-energy electron microscopy (LEEM) to study graphene growth on Ru from vapor deposited and segregated C. We measured the growth rates as a function of the concentration of the mobile, gas-like C adatoms that surround the graphene. We showed that the growth rates on the Ru(0001) surface are not limited by the surface diffusion of C adatoms and that an energy barrier exists for attaching single adatoms to the edges of graphene sheets [26; 27]. Furthermore, graphene crystal growth is not in the linear, close-to-equilibrium regime generally observed for the growth of metal and semiconductor films on metal surfaces [28-30]. A possible explanation for a large energy barrier to graphene growth on Ru is that C can be strongly bound to both Ru and to graphene but not to both simultaneously (i.e., graphene is relatively weakly bound to the metal). Thus, C motion from the metal to the graphene requires an intermediate, energetically costly state. Indeed, our analysis of the nonlinearity of the growth rate as a function of adatom density suggested that this intermediate state involves clusters of about five C atoms [26]. Presumably these clusters allow the attaching C atoms to remain bonded to other atoms to the best extent during the entire attachment (or detachment) process. There are two limiting scenarios for this cluster-addition kinetics, which our previous work did not distinguish. In one limit, the C clusters are constantly created and destroyed as thermal excitations everywhere on the Ru terraces, and have no energy barrier to attaching to the graphene step edge as they arrive there. In this case the activation energy of growth (measured to be 2 eV [26]) would correspond to the C-cluster formation energy. In the second limit, clusters are only formed during an attachment event. In this scenario, the activation energy depends on the structure of the graphene edge.

To establish the generality of the cluster-addition mechanism and learn where attaching clusters are formed, we have extended our previous approach in two ways: we analyzed the growth kinetics using ethylene molecules as the C source, and, we studied graphene growth on a different substrate, Ir(111). In graphene growth from ethylene on Ru(0001), we found exactly the same nonlinear dependence of graphene growth rate on C adatom concentration as for growth from elemental C. This means hydrogen from ethylene changes neither the energetics of the C clusters nor the bonding of the graphene step edges.

That said, the rate at which ethylene decomposes to form adatoms *does* affect the overall graphene growth rate. For example, because ethylene does not stick or decompose significantly on the graphene, the growth rate decreases as the surface coverage of graphene increases.

Ir(111) is particularly interesting substrate for graphene, we have found that it supports growth of discrete crystallographic orientations. The majority orientation, described in detail in the literature [9; 10; 20], exhibits nonlinear growth kinetics similar to graphene on Ru(0001). The observation of nonlinear growth kinetics suggests that the difficulty of monomer attachment to step edges is not limited to Ru. This similarity in growth kinetics occurs even though density functional theory (DFT) calculations and experiment show that C adatoms are less strongly bound to Ir than to Ru.

The minority orientations of graphene on Ir grow markedly faster than the majority orientation at the same concentration of C adatoms. This orientation dependence is evidence against the notion that cluster formation on terraces is rate-limiting for graphene growth. It implies instead that the growth rate is governed by C-cluster attachment to the edges of the graphene flakes.

A clear difference between graphene growth on Ir and Ru is the relationship to substrate steps. Graphene grows almost exclusively down the staircase of Ru steps [26; 8]. In contrast, here we find graphene grows both up and down the Ir staircase, consistent with the inferences of Ref. [10].

Lastly, we discuss the nucleation process. We show that nucleation occurs at the same adatom concentration independent of whether elemental carbon or ethylene is the C-atom source, and interpret this concentration in terms of the critical nucleus size.

**2. Experimental Methods**

All LEEM experiments were conducted in an ultrahigh vacuum system with base pressure $\sim 1\times 10^{-10}$ Torr. The Ru(0001) and Ir(111) crystals were depleted of bulk C by cycles of annealing for several hours in UHV at ~1000 K, followed by oxygen exposure at pressure $1\times 10^{-8}$ Torr and temperature ~800 K, then, flashing to ~1660 K in vacuum. Graphene was grown by exposing the Ru surface at 750 to 1000 K and the Ir surface at 850 K to 1320 K to ethylene gas of 99.999% purity. Over these temperature ranges C diffusion into the crystal bulk and C surface segregation from the bulk are negligible. Ethylene pressure is measured with an ionization gauge and reported without correction for gauge sensitivity. However, a gas correction factor of 2.6 was used to calculate the ethylene sticking probabilities. Temperatures were measured by type-C thermocouples spot-welded to the sides of the crystals. The structure of the graphene was analyzed from low-energy electron diffraction (LEED) patterns obtained from single islands by using apertures to limit the size of the incident electron beam.

During ethylene deposition, the concentration of C adatoms was measured using the change in the intensity of electrons reflected from the surface, as previously described [31; 26]. By measuring the reflectivity change from LEEM images, we can determine the C adatom concentration adjacent to growing graphene islands. To apply the technique to graphene growth from ethylene, we first show that the choice of C source (elemental C or ethylene) does not affect our previous calibration for Ru(0001) [26]. For Ir(111), we describe our calibration procedure.

Figure 1 (a) compares the electron reflectivity as a function of electron energy for Ru(0001) in two conditions, clean and after exposure to sufficient ethylene to nucleate graphene islands. In the latter case, we use the power of our imaging approach to measure the reflectivity from the area between the graphene islands shown in Figure 1 (b). The decrease in electron reflectivity as a function of electron energy resulting from ethylene exposure is identical to that produced by depositing elemental C (Ref. [26], Figure 1(a)) or segregating C from the bulk of Ru (Ref. [27], Figure 1). This agreement shows that the reflectivity change is caused by mobile C adatoms and that hydrogen does not contribute to electron reflectivity changes. The latter conclusion is consistent with previous observations that ethylene molecules completely dissociate on Ru(0001) and that the hydrogen desorbs at temperatures well below our graphene growth temperatures [32; 33]. To confirm further that electron reflectivity is unaffected by hydrogen, we established that exposing a clean Ru(0001) surface at 940 K to hydrogen pressures up to $1\times10^{-6}$ Torr had no measurable effect on electron reflectivity. As in our previous study, we measure reflectivity changes using an incident electron energy of 3.7 eV, which is near the energy of maximum sensitivity (see Fig. 1 (a)). In section 3.3, we provide further evidence that C adatoms originating from ethylene and vapor-deposited C affect electron reflectivity in exactly the same manner. We found that the growth rate could affect the crystalline quality of the graphene. For example, "slow" graphene growth at 890 K with ethylene pressure below $\sim3\times10^{-9}$ Torr yields high-quality epitaxial graphene islands on Ru(0001), as established by the sharp superstructure diffraction spots up to seventh order in the LEED pattern of Figure 1 (c). "Fast" growth with ethylene pressure above $\sim1\times10^{-8}$ Torr produces less-ordered graphene with a diffuse LEED pattern. Annealing to 1200 K, however, ordered the graphene yielding a sharp diffraction pattern.

For the Ir(111) surface (see Figure 2 (a)), the difference in reflectivity between a clean and a C-adatom-covered surface is smaller than for Ru(0001). We measure reflectivity changes at 17.0 eV, chosen as a compromise between having a large change in reflectivity from C adatoms and high reflected intensity from graphene-free regions. To determine absolute concentrations of C adatoms on Ir(111), we first established that C deposition at constant rate from an elemental C source decreased the reflectivity linearly in time. The reflectivity change is then linearly proportional to the adatom concentration. The reflectivity change was calibrated to determine absolute C adatom concentration using the previously described [26] procedure. Specifically, C was deposited at a constant rate from a graphite rod heated by electron-beam bombardment until the surface was partially covered by graphene islands. The absolute deposition rate was determined by measuring the final fractional surface coverage of graphene islands by averaging the coverages of 10 LEEM images from different areas. After correcting self-consistently for the small fraction of the deposited C in the adatom gas, we find that the C adatom coverage on Ir(111) equals $0.189 \times (I_0 - I(t))/I_0$ ML. Here $I_0$ is the initial reflected intensity, measured from local regions of LEEM images using 17.0 eV electrons, and *I(t)* is the intensity at time *t*. 1 ML is defined as the areal density of graphene on Ir(111). In comparison, the calibration for Ru using 3.7 eV electrons is $0.223 \times (I_0 - I(t))/I_0$ ML.

**3. Results and discussion**

*3.1 Establishing that ethylene does not decompose on graphene*

In our experiments, graphene was nucleated and grown via thermal decomposition of ethylene. Adsorbed ethylene molecules are well-known to dissociate completely on Ru(0001) upon annealing to 720 K, producing surface C and hydrogen that recombines and desorbs from the surface [32; 33]. During ethylene deposition, we find that the rate at which graphene covers the Ru substrate depends on the fractional coverage of graphene. Let $\mathcal{A}$ be the fraction of surface area covered by graphene. The time dependence of the fractional graphene coverage $\mathcal{A}$ is demonstrated in the top insert of Figure 3. Figure 3 shows the time rate of change of surface coverage, $d\mathcal{A}/dt$, going from a graphene-free surface to being completely covered with graphene. We find $d\mathcal{A}/dt \propto (1-\mathcal{A})$, where $1-\mathcal{A}$ is the fraction of the surface that is uncovered (bare Ru). Thus, the rate at which graphene grows is proportional to the amount of exposed Ru. This means that ethylene sticks to and decomposes on uncovered Ru only (see Figure 3 (bottom insert)).

### *3.2 Sticking probability of ethylene on Ru(0001) at 1020 K*

Knowing C monomer concentrations, we can calculate the ethylene sticking probability, which is the ratio of the flux of incident ethylene molecules to the rate of change of adsorbed C adatoms. The ethylene sticking probability $s$ is:

$$s = \frac{dn_{ads}/dt}{2 dn/dt} = \frac{\rho dc/dt}{2p/\sqrt{2\pi m k_B T}}, \tag{1}$$

where $n_{ads}$ and $n$ are the numbers of adsorbed C monomers and incident ethylene molecules, respectively; $c$ is the C monomer concentration in monolayers (see section 2); $\rho$ is density of C atoms in graphene; $p$ is the ethylene pressure measured using a pressure gauge sensitivity factor of 2.6 (the factor for $C_2H_6$ – the closest molecule to $C_2H_4$ found in the literature [34]), $m$ is the molecular weight of ethylene, $k_B$ is the Boltzmann constant and $T$ is the ethylene temperature (300 K). The factor of 2 in denominator comes from the assumption that every ethylene molecule yields two C adatoms. Figure 4 shows the ethylene sticking probability for C monomer concentrations during ethylene deposition at a constant ethylene pressure of $5\times10^{-9}$ Torr. The maximum sticking probability is 0.190 at C monomer coverage 0.005 ML. At higher coverage, the ethylene sticking probability decreases roughly linearly, reaching a value of 0.04 when graphene nucleation occurred ($c^{nucl}$=0.035 ML). The maximum sticking probability is close to the sticking probability of $C_2H_4$ on Pd(110) at 800 K measured with temperature-programmed adsorption by Bowker et al. [35] and given in an isothermal uptake curve. A similar value of the sticking probability of $C_2H_6$ on Pt(110) was obtained by Harris et al. [36] using a supersonic molecular beam and a temperature-programmed reaction technique for ~0.05 ML coverages at 600 K. While the accuracy of this sticking probability measurement is limited by knowledge of the absolute $C_2H_4$ pressure, the result demonstrates the ability of the electron reflectivity technique to measure quantities such as sticking probabilities.

### *3.3 C adatom supersaturation and nonlinear graphene growth on Ru(0001)*

Having described where ethylene decomposition occurs (section 3.1) and how the ethylene sticking probability changes with C adatom coverage (section 3.2), we now analyze the relationship between the graphene growth rate and the C adatom concentration resulting from ethylene decomposition. Figure 5 shows how the C adatom concentration changes during ethylene exposure. The concentration initially increases, reaching a maximum and then decreases. Inspection of the images establishes that the C adatom concentration begins to decrease, even though the ethylene pressure is held constant, when graphene islands nucleate. After ethylene exposure is stopped, the C adatom concentration decreases very slowly with time until it comes into equilibrium with the graphene islands at concentration $c^{eq}$ = 0.017 ML. The ratio of the real-time concentration, $c$, to $c^{eq}$ is about 2 at nucleation ($c^{nucl}$ = 0.035 ML) and between 1.5 and 2 during growth. Thus, we find that a large supersaturation is needed to nucleate and grow graphene from ethylene, in qualitative agreement with our observations for graphene growth on Ru fed by elemental C. We next show that the detailed dependence of growth rate on C adatom concentration is quantitatively the same when ethylene and elemental C are used as sources.

As previously discussed [26], the essential measure of crystal growth is the velocity at which the edge of a graphene sheet advances as a function of the surrounding adatom supersaturation, $c/c^{eq}$. As before, we measure velocity $v$ by determining the area $A$ and perimeter $P$ of discrete islands: $v = P^{-1}dA/dt$. Figure 6 shows that the graphene growth velocities for ethylene deposition and vapor-deposited C are absolutely identical functions of the mobile C adatom concentration. The growth rates are highly nonlinear with supersaturation, which is very unusual for crystal growth [37]. The crystal growth velocity, which is the difference between attachment and detachment rates of a growth species, usually changes linearly with monomer concentration. The growth velocity is well-described by the expression:

$$v \propto \left[\left(\frac{c}{c^{eq}}\right)^n - 1\right] \qquad (2)$$

with $n \approx 5$. In our model of cluster-addition kinetics, $n$ is the number of C atoms in the growth species that attaches to the graphene edge [26].

Since we observe the same nonlinear dependence of growth velocity $v$ on supersaturation for C segregating from the Ru bulk, we conclude that the graphene growth mechanism is independent of the C adatom source, i.e., vapor-deposited [26], segregated [27], or dissociated from ethylene. In all cases, the large energetic barrier for C adatom attachment to graphene leads to sluggish equilibration of the adatoms with the graphene and flat concentration profiles of the adatoms across the surface. Because the adatom concentration is uniform, the edge velocity is independent of island size and environment [26]. The graphene growth rate is completely determined by the C adatom concentration, independent of the source. However, the rate of adatom generation, and, thus, the global growth rate, does depend on the C source. The rate at which C adatoms are generated from ethylene decomposition depends upon the fractional coverage of graphene (section 3.1 and Fig. 3) and somewhat upon the C adatom coverage (section 3.2 and Fig. 4). Similarly, the rate of C segregation can be limited by bulk diffusion [27]. We next consider the role of the substrate in the nonlinear growth kinetics by studying graphene growth on Ir(111).

### *3.4 Growth kinetics of the majority orientation of graphene on Ir(111)*

We have found that graphene forms at least four rotational variants on the Ir(111) surface. The most abundant, the "majority" orientation, has been studied in detail [9; 20; 10]. In this section, we explore its growth and find essentially the same nonlinear kinetics as for graphene on Ru. In the next section, we show that the minority rotational variants have markedly different kinetics.

Figure 7 shows electron reflectivity (on the left axis) and calibrated C adatom concentration (on the right axis) as a function of time during ethylene exposure at 1100 K. Starting from the clean substrate, the C adatom concentration built up during the exposure. The adatom concentration began to decrease once graphene islands nucleated. Only graphene islands of the majority orientation nucleated within the field-of-view of the experiment (see the image in Fig. 7). After ethylene exposure was stopped, the adatom concentration very slowly came into equilibrium with graphene, with $c^{eq} = 0.007$ ML. Similar to graphene growth on Ru(0001), large supersaturations are needed to nucleate and grow these graphene islands of majority orientation.

For example, $c^{nucl}$ = 0.013 ML, about twice the equilibrium concentration. The smaller change in reflectivity for a given C adatom concentration (see section 2) on Ir(111) compared to Ru(0001) and the existence of the other rotational variants of graphene currently limits our ability to measure the adatom formation energy and the activation energy for growth to the same precision as for C on Ru.

Comparing Figs. 5 and 7 shows that the equilibrium concentrations of adatoms are lower on Ir than on Ru. This observation suggests that the enthalpy $E_{Ir}^f$ of forming a C monomer on Ir(111) from graphene is larger than the formation enthalpy $E_{Ru}^f$ on Ru(0001). We can estimate $E_{Ir}^f$ using the equilibrium concentration for Ru(0001) at 1020 K and the relationship:

$$\frac{c_{Ru}^{eq}(T_{Ru}=1020K)}{c_{Ir}^{eq}(T_{Ir}=1100K)} = \frac{0.018ML}{0.007ML} = e^{-\left(\frac{E_{Ru}^f}{kT_{Ru}} - \frac{E_{Ir}^f}{kT_{Ir}}\right)}. \quad (3)$$

Equation (3) assumes that the formation entropies of the monomer are the same on the two substrates. Solving for $E_{Ir}^f$ using the experimental and calculated value of $E_{Ru}^f \approx 0.3$ eV [26] gives $E_{Ir}^f \approx 0.4$ eV. To determine if this result is reasonable (and, thus, again test the validity of our claim that the reflectivity changes we observe are attributable to C adatoms), we have used DFT calculations in the Perdew-Wang '91, Generalized Gradient Approximation (PW91-GGA) [38; 39] to estimate $E^f$. We indeed find $E^f$ is significantly higher on Ir, consistent with experiment: it equals roughly 0.5 eV on Ir(111), but 0.3 eV on Ru(0001). Details of the calculations yielding these results are provided in the Appendix.

The high supersaturations needed to nucleate and grow graphene on Ir, the extremely sluggish equilibration between the adatoms and the graphene and the flat concentration profiles of the C adatoms across the surface show that an energy barrier exists to attaching C adatoms to graphene step edges. Thus, growth of the majority orientation of graphene on Ir is qualitatively similar to graphene growth on Ru. We note that graphene on Ru and the majority orientation on Ir both have exactly the same crystallographic orientation of the graphene sheets relative to the close-packed directions of the substrates [10; 12]. (The two systems do differ in the size of the unit cell and the amount of corrugation [24].) Thus, on both metals an energy barrier exists to attaching individual C atoms to this orientation of graphene relative to the substrate.

Examining how the growth rate of the majority orientation of graphene on Ir depends on C adatom concentration establishes that the growth mechanism is the same as for Ru. Figure 8 shows that the rate at which a graphene step edge advances on Ir(111) is also a highly nonlinear function of C adatom concentration. In fact, the data is well-described by the model of cluster-addition kinetics, eqn. 2, with a cluster size *n* of 5 C atoms. More experimental work is required, however, to establish the order of the growth kinetics with the same level of confidence as done for graphene growth on Ru. Overall, the large similarities of the growth kinetics of the majority orientation of graphene on Ir(111) and graphene on Ru(0001) suggest that the growth mechanism is the same. Our interpretation is again that clusters of about 5 C atoms add to graphene, rather than the abundant monomers adding individually.

### *3.5 Growth kinetics of a minority orientation of graphene on Ir(111)*

Elsewhere, we report detailed characterization of the atomic structure of the orientational variants using LEEM, LEED, and scanning tunneling microscopy [40]. The other variants are also "moiré-like" structures of a graphene sheet lying on the imperfectly lattice-matched substrate. However, the graphene sheets in the "minority" variants are rotated by approximately 14°, 18.5° and 30° relative to the majority orientation.[1] Thus, graphene/Ir(111) is similar to the graphene/Pt(111) system, where several rotational variants also coexist [11; 25]. A strong contrast exists between the majority rotational variant and the minority variants in bright-field LEEM images, allowing their formation and growth to be easily monitored [40]. Typically we observe that islands of the majority orientation nucleate first. Minority variants then nucleate at the perimeter of the growing islands. Figure 9 makes the point. The image at 400 s shows an island with majority orientation shortly after a minority-type island nucleated on its upper-right perimeter. At the time of the minority phase nucleation, the adatom concentration has dropped by ~30% from the value at which the majority orientation nucleated. The minority orientation appears bright compared to the majority orientation. Even though the two graphene orientations experience the same adatom concentration, the minority (bright) island grows much more rapidly, as can be concluded directly by inspecting the images. The left axis of the plot in Fig. 9 shows that the difference in growth rates, as defined in section 3.3, can be greater than a factor of 4. The growth rates of both graphene islands dropped with time because the ethylene pressure was decreased (right axis of the plot in Figure 9 (a))[2]. Similar results were observed for each island of minority orientation, despite their differing orientations.

Nucleation of the minority phases on the edges of the majority-phase graphene at relatively small adatom coverages suggests that nucleation of the minority phases is heterogeneous. That is, nucleation of the minority phases is initiated by rare defects in the majority phase, such as the edge dislocations described in Ref. [9].

---

[1] These graphene rotational angles are much larger than those reported in ref. [9] Coraux J, N'Diaye A T, Busse C, and Michely T 2008 Structural coherency of graphene on Ir(111) *Nano Letters* **8** 565-70 for the majority orientation and should not be confused with the large rotations of the moiré unit cell that small rotations can cause.

Growth velocities of the two rotational variants in Fig. 9 might differ for two reasons. The first is that the majority and minority phases are growing from different supersaturations because the different phases have different equilibrium adatom concentrations. Alternatively, the growth mechanisms are different. The first (and simpler) possibility would occur were the binding energy of the graphene sheet of the minority orientation to the Ir lower than the binding energy of the majority orientation. That would affect the equilibrium adatom concentration because adatom formation energies are measured relative to the binding of C in the graphene (see Appendix). Owing to the higher adatom formation energy, the adatom concentration in equilibrium with the minority graphene would then be lower than that in equilibrium with the majority graphene. Thus, the minority phase would be farther from equilibrium with the surrounding adatom gas and grow more quickly, assuming the kinetic mechanisms to be the same (c.f. Eq. (2)).

This argument seems physically improbable because it would require the rotated moire's to be considerably more strongly bound to the substrate, presumably by the formation of chemical bonds. However, the majority orientation has been found to be weakly van-der-Waals bonded to the underlying Ir(111) surface [20]. Since all moire's contain the same relative numbers of carbon atoms near the various binding sites of Ir, it is unclear why rotation would the nature of this binding to change. Beyond theoretical notions, experiment provided direct support for rejecting this scenario, namely, the islands of the majority and minority orientations stopped growing simultaneously when the adatom concentration dropped (see Fig. 9).

The second reason why the growth velocities of the two rotational variants could be different is that the C attachment kinetics of the two phases are different. Because the two orientations experience the same adatom gas, this in turn implies that the graphene step edges have an influence on the cluster-attachment kinetics. We next develop this proposal by comparing and contrasting the growth kinetics of the only orientation of graphene on Ru and the different orientations of graphene on Ir.

---

[2] The difference in growth rates occurs even though the island with minority orientation can largely grow only in the slower "uphill" direction. See section 3.6.

Graphene on Ru and majority type of graphene on Ir have the same orientation of the graphene sheets relative to the high-symmetry directions of the substrates [12; 10]. This graphene orientation has a similar nonlinear dependence of growth velocity on C adatom supersaturation on both substrates. The kinetics suggest a mechanism where growth occurs by the addition of C clusters of about 5 C atoms, independent of the substrate. In contrast, the rotated orientations of graphene on Ir grow markedly faster at the same C adatom supersaturation. The variants on Ir that are rotated relative to both graphene on Ru and the majority type of graphene on Ir likely differ in how the edges of the graphene sheets are terminated and bonded to the substrate. That is, the different orientations likely have different arrangements of C atoms (e.g., "zigzag" vs. "armchair") and/or different bindings to the substrate. Thus, we are led to the conclusion that attachment of the cluster to the edge of the graphene sheet is the slow step in the nonlinear kinetics (cluster-addition mechanism), rather than the formation of the clusters. (It is thus possible that the reported preferred orientation of graphene edges [9] are due to kinetic selection rather than energy minimization.)

### *3.6 Directionality of graphene growth on Ru(0001) and Ir(111)*

In this section, we discuss the rates at which graphene sheets grow on Ru and Ir relative to directions that go up or down substrate steps. As is common for single crystals of metals that are precisely polished, our Ru and Ir crystals have reasonably well-defined "staircases" of surface steps. That is, most of each surface is covered by an array of atomic substrate steps with a well-defined direction that runs up (or down) the staircase. Figure 10 shows a sequence of LEEM images taken during graphene growth on Ru from ethylene at 960 K. The graphene island does not grow "uphill" by crossing over the Ru step edge where it nucleated, consistent with previous observations of Sutter et al. [8] and ourselves [26]. Instead, the graphene island grows by extending itself in the direction down the staircase of the Ru steps. The island also grows along the Ru step edge where it nucleated, reaching a width of about 7 μm in this direction. Thus, the smooth arc on the island's left side is defined by the Ru step edge. The exclusively "downhill" growth of graphene on Ru occurs above about 850 K using ethylene.

In contrast to the downhill growth of graphene on Ru(0001) surface, the majority and minority rotational variants of graphene on Ir(111) grow in both downhill and in uphill directions. (See also the example in Fig. 9.) Figure 11 shows the growth of the majority graphene orientation during ethylene deposition at 1200 K. While the graphene island advances in all directions, there is still a preference for growing down the Ir steps. Nonetheless, the growth mechanism, as inferred from the growth kinetics, is insensitive to the relative amount of uphill and downhill growth.

*3.7 Critical nucleus size for graphene nucleation on Ru(0001)*

As Figure 5 shows, nucleation on Ru(0001) is observed at a specific C-monomer concentration. Figure 12 shows that the adatom concentrations at nucleation are the same for vapor-deposited C and ethylene exposure. Thus, the decomposition of ethylene influences the initial nucleation of graphene.

Figure 12 also compares the temperature dependence of this concentration to the concentration in equilibrium with graphene. Interestingly, $c^{nucl}$ is about twice $c^{eq}$ for all temperatures. As we will now argue, this result is consistent with classical nucleation theory if the critical nucleus size is roughly constant. This independence of critical nucleus size on temperature gives insight into what determines the density of nucleated islands.

Classically, the graphene nucleation rate for a given supersaturated monomer concentration is proportional to the probability of the formation of an island of the critical size [41]:

$$R^{nucl} \propto e^{-G(n^*)/kT}. \tag{4}$$

In Eqn. (4), $G(n)$ is the free energy of an island with $n$ atoms, and $n^*$ is the minimum size of an island that will grow from the monomer environment. Figure 12 shows that this rate becomes appreciable at all temperatures when $c \approx 2c_{eq}$, i.e., when the carbon chemical potential with respect to graphene is

$$\mu = kT\ln(c/c_{eq}) = kT\ln 2. \tag{5}$$

The dependence of $G(n)$ on $\mu$ is given by

$$G(n) = E(n) - \mu n, \tag{6}$$

where $E(n)$ is the energy difference between $n$ carbon atoms in an island or cluster and a graphene crystal. Substituting Eqs. (5) and (6) into Eq. (4) gives a nucleation rate of

$$R^{nucl} \propto 2\,e^{n^*} \exp(-E(n^*)/kT) \tag{7}$$

This rate, at which nucleation is large enough to be observable depending only on the observer's patience, must be approximately independent of temperature. The simplest scenario for this is that $n^*$ does vary with temperature and $E(n^*)$ is not much larger than $kT \approx 0.1\text{eV}$. The latter requirement implies that the critical island does not contain many broken C-C bonds, which cost several eV. This in turn suggests that the graphene edges are rebonded or that there is significant binding to the metal.

Despite the fact that the critical nucleus size does not appear to change much with temperature, we do observe that the number of nucleated islands decreases rapidly with increasing temperature. For growth from ethylene, as well as C-vapor deposition, the epitaxial graphene island density depends strongly on the substrate temperature: for an ethylene pressure of $5 \times 10^{-9}$ Torr, the nucleation density on Ru(0001) is relatively high at 740 K and sparse at 1070 K, yielding nuclei separated by up to 50 μm. We attribute this difference to the strong temperature dependence of the rate at which the monomer concentration equilibrates after islands are nucleated. At low temperature, the concentration remains large even after islands start to nucleate because the attachment rate at island boundary sinks is so small (with an activation energy of 2 eV [26]). Thus, there is more time for more islands to nucleate at lower temperatures. Notice in the normal theory of nucleation the equilibration time depends on diffusion rates [41] – in our case it is determined by the monomer attachment barrier.

**4. Summary and conclusions**

In summary, we have probed how graphene grows on metal surfaces by contrasting growth from a hydrocarbon, ethylene, with growth from vapor-deposited elemental C and segregating carbon. We also compare two substrates, Ru(0001) and Ir(111). For growth temperatures that lead to high-quality epitaxial graphene, we find that the growth mechanism does not change with the C source. This insensitivity occurs because, both the nucleation and the growth occurs from mobile C atoms on the surface. However, the rate at which adatoms are generated, and, thus, the overall growth rate does depend on the C source since ethylene only decomposes on bare substrate and the ethylene sticking probability decreases with C adatom concentration.

Our analysis of graphene growth rates as a function of adatom concentration for the graphene orientation on Ir(111) previously described in detail [10] shows nonlinear growth kinetics that are similar to Ru. This suggests the cluster addition mechanism of growth is not specific to Ru. However, Ir(111) provided an unexpected insight into growth mechanisms in that graphene forms in three orientations rotated from the previously described majority phase. The rotated orientations grow significantly faster at the same C adatom concentration. The observation that the growth rate changes with the orientation of the graphene sheet suggests that the arrangement of the C atoms at the edges of the graphene sheets and the bonding of the edge atoms to the substrate play important roles in the growth mechanism. Specifically, the rate-limiting step of the nonlinear kinetics is likely the attachment or formation of the C cluster at the edges of the graphene sheets, not the rate at which clusters are created on terraces.

**Acknowledgments**

This work was supported by the Office of Basic Energy Sciences, Division of Materials Sciences and Engineering of the US DOE under Contract No. DE-AC04-94AL85000.

**Appendix**: *Methods of calculating C adatom formation energies on Ir(111) and Ru(0001)*

We describe here the density functional theory (DFT) calculations in the Perdew-Wang '91, Generalized Gradient Approximation (PW91-GGA) [38; 39] used to calculate C adatom formation energies (see section 3.4). We computed PW91-GGA formation energies of 0.48 eV for a C-adatom on Ir(111), and 0.31 eV on Ru(0001), by subtracting the C binding energy, in its preferred site on each metal surface, from the binding energy per C atom in an extended graphene sheet adsorbed on the same metal. The energy evaluations were performed using VASP [42; 43] with electron–core interactions treated in the projector augmented wave approximation [44; 45]. Electronic convergence was accelerated by means of Methfessel-Paxton smearing of the Fermi level (width = 0.2 eV) [46]. The Neugebauer-Scheffler method [47] was used to cancel unphysical fields resulting from the differences imposed on the upper and lower slab surfaces. Lattice constants were fixed by optimization of the bulk metal crystals. The basis cutoff was set to 700 eV.

To quantify the numerical accuracy of the computed binding energies (cf., Table 1 for the Ir(111) case) we compared results for different supercells, $3 \times 2\sqrt{3}$, and $2\sqrt{3} \times 2\sqrt{3} - R30°$, slab thicknesses from 7 to 9 layers (in each case, fixing in bulk relative positions the atoms of lowest three layers), and several refinements of the surface brillouin zone (SBZ) sample. The supercell comparison showed that with a nearest C-C distance of 8.24 Å in the repeats of the $3 \times 2\sqrt{3}$ case, the C-C interaction contributes at the level of 0.01 eV. We found for the $2\sqrt{3} \times 2\sqrt{3} - R30°$ supercell that a 6×6, equal-spaced SBZ sample of k-vectors provides 0.01 eV accuracy. Lastly, the ad-C binding energy versus slab thickness is still changing by ~0.05 eV between 7 and 8, and between 8 and 9 Ir(111) slab layers, suggestive of a quantum size effect [48]. We plan to investigate this effect in detail. In the meantime, the formation energy estimate quoted above for Ir corresponds to C adsorption on a 10-layer slab.

It should be noted that, as in the case of Ir self-adsorption on Ir(111) [49] the preferred binding site for the isolated C adatom was the hcp, "stacking fault" and not the fcc, or "lattice continuation" site, with the preference amounting to 0.24 eV. In self-adsorption on Ir(111), such behavior has been attributed to a band-structure effect [50].

To complete the calculation of the C formation energy, one must compute the binding of a C atom in graphene adsorbed on Ir(111). Within the PW91-GGA, however, this binding is virtually entirely the cohesive energy of an isolated graphene layer [51]. It was evaluated as in [26]. Therein, there is also a discussion of the binding energy calculation for C/Ru(0001).

| supercell | Ir lyrs | SBZ sample | C site | $E_{form}$ |
|---|---|---|---|---|
| 3×2√3 | 7 | 8×8 | hcp | 0.54 eV |
| 3×2√3 | 8 | 8×8 | hcp | 0.49 eV |
| 3×2√3 | 9 | 8×8 | hcp | 0.45 eV |
| | | | | |
| 2√3×2√3 | 7 | 6×6 | hcp | 0.56 eV |
| 2√3×2√3 | 7 | 9×9 | hcp | 0.55 eV |
| 2√3×2√3 | 8 | 6×6 | hcp | 0.50 eV |
| 2√3×2√3 | 9 | 6×6 | hcp | 0.45 eV |
| 2√3×2√3 | 10 | 6×6 | hcp | 0.48 eV |
| | | | | |
| 2√3×2√3 | 7 | 6×6 | fcc | 0.80 eV |
| 2√3×2√3 | 7 | 9×9 | fcc | 0.79 eV |

Table 1. Numerical accuracy tests for the PW91 formation energy of a C monomer on Ir(111). The 3×2√3 supercell implies a nearest C-C distance of 8.24Å. With a $2\sqrt{3} \times 2\sqrt{3}$ periodicity, the nearest C atoms are separated by 9.52Å. As in the case of Ir ad-monomers on Ir(111), the stacking fault (i.e., hcp) site is favored over the lattice continuation (i.e., fcc) binding site.

**Figures**

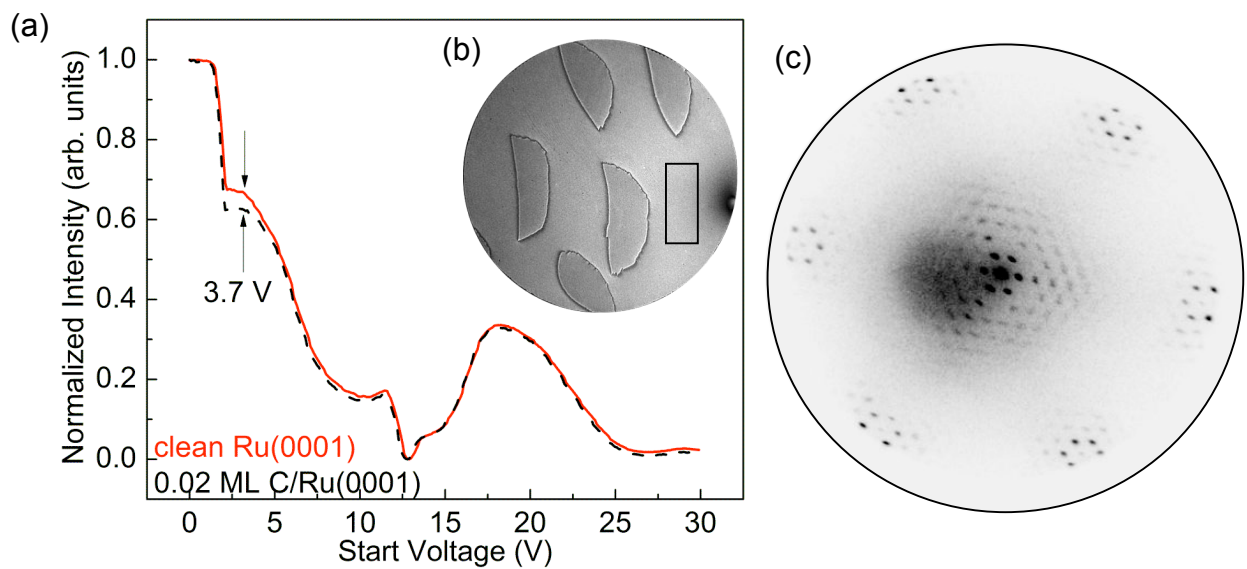

Figure 1. (a) Electron reflectivity as a function of incident electron energy for clean Ru(0001) (solid red line) and covered with 0.020 ML of C-adatoms (dashed black line) in equilibrium with 0.2 ML of graphene at 980 K after $C_2H_4$ deposition at 1020 K; (b) LEEM image (20 μm field-of-view) with 5 graphene islands. The black box indicates the graphene-free region analyzed in (a). (c) LEED image taken at electron energy 32 eV and temperature 300 K from a single graphene island grown at 890 K and $C_2H_4$ pressure $3\times10^{-9}$ Torr.

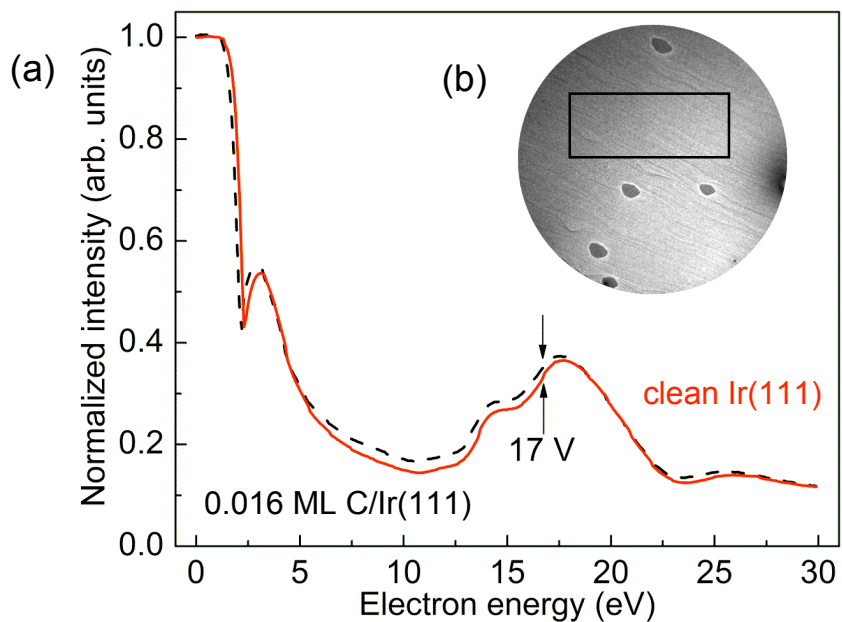

Figure 2. (a) Electron reflectivity dependence on incident electron energy for clean Ir(111) (solid red line) and covered with 0.016 ML of C adatoms (dashed black line) in equilibrium with graphene after $C_2H_4$ deposition at 1100 K; (b) LEEM image (46 μm field-of-view) with a graphene island. The black box indicates the graphene-free region analyzed in (a).

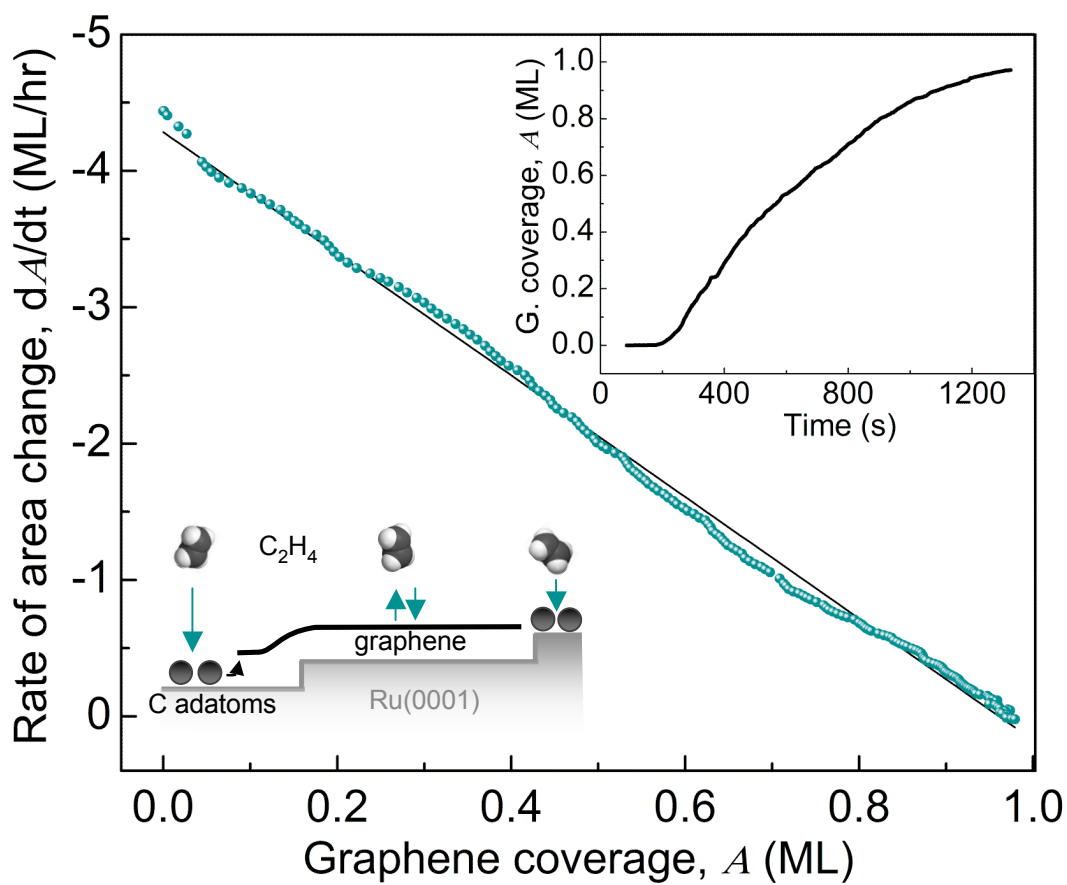

Figure 3. Rate of change of graphene coverage, $d\mathcal{A}/dt$, as a function of the fraction of the surface covered by graphene $\mathcal{A}$, during ethylene deposition at pressure $3\times10^{-8}$ Torr at 1020K. Inserts show time dependence of graphene coverage $\mathcal{A}$ and schematic representation of ethylene deposition: ethylene molecules decompose on the graphene-free Ru surface producing carbon adatoms. $C_2H_4$ molecules do not dissociate on the graphene-covered surface.

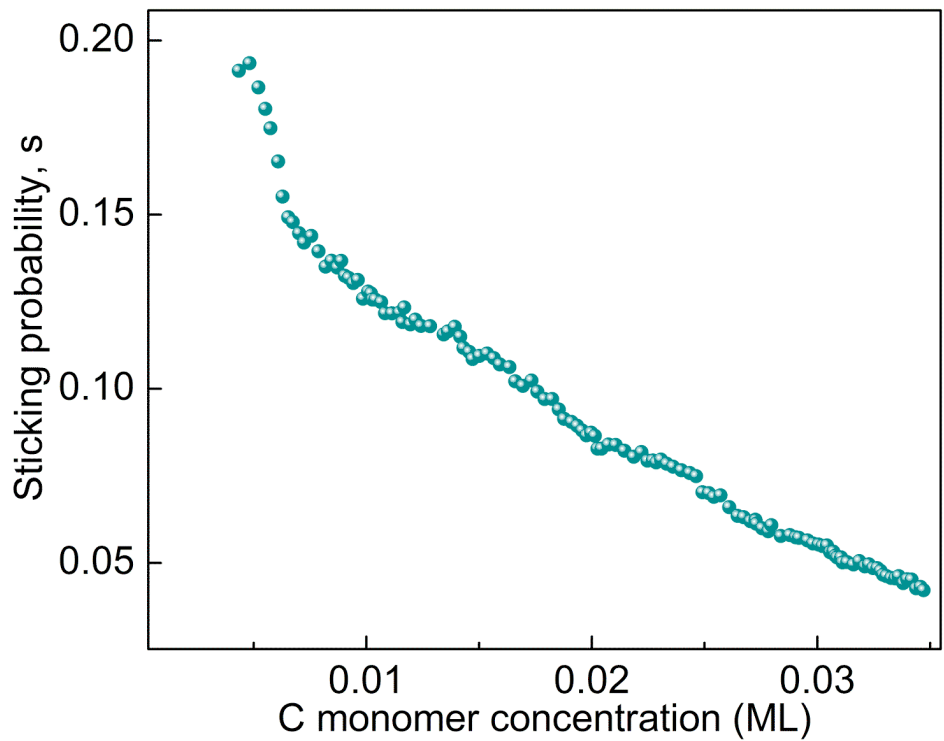

Figure 4. Sticking probability of ethylene to Ru(0001) surface at 1020 K, measured using the change in electron reflectivity during $C_2H_4$ deposition at constant pressure $5\times10^{-9}$ Torr.

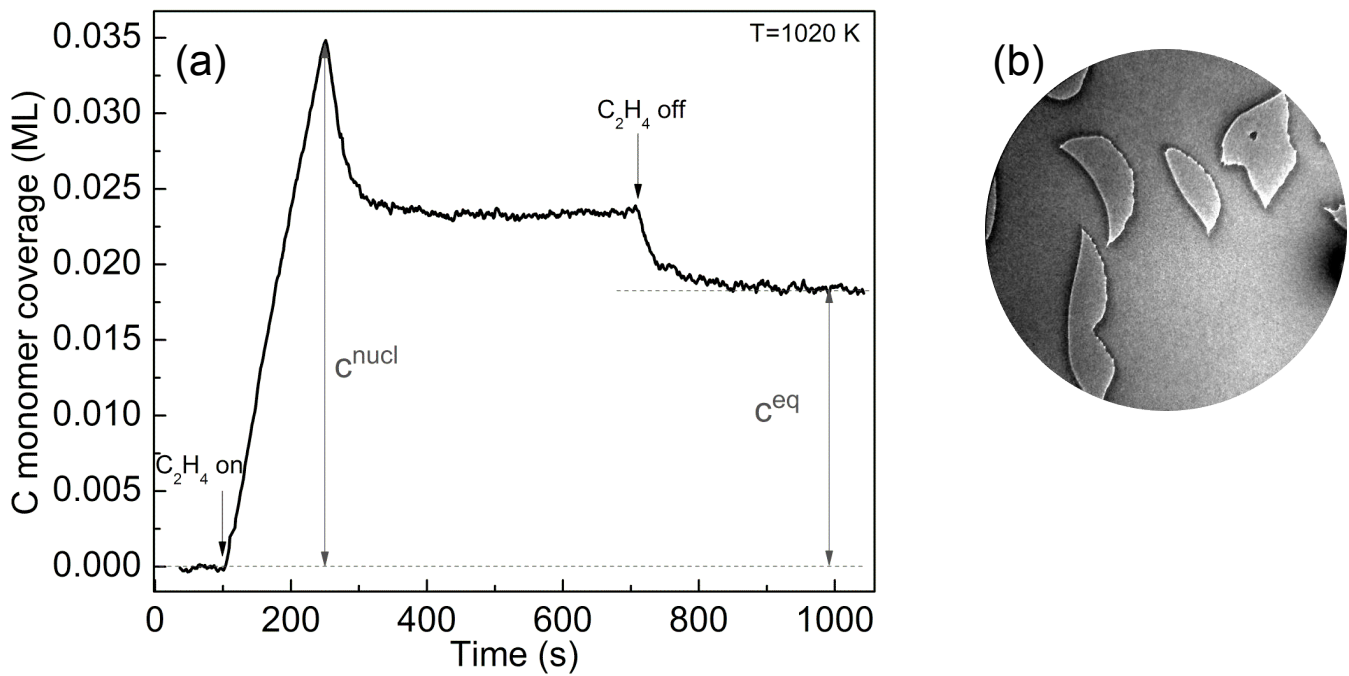

Figure 5. (a) The C monomer concentration on Ru(0001) in equilibrium with graphene measured from electron reflectivity during ethylene deposition at constant pressure $3\times10^{-9}$ Torr at 1020 K; (b) LEEM image (20 μm field-of-view) taken at 1000 s. Concentrations $c^{nucl}$ and $c^{eq}$ are the C monomer concentrations needed to nucleated graphene and be in equilibrium with graphene islands, respectively.

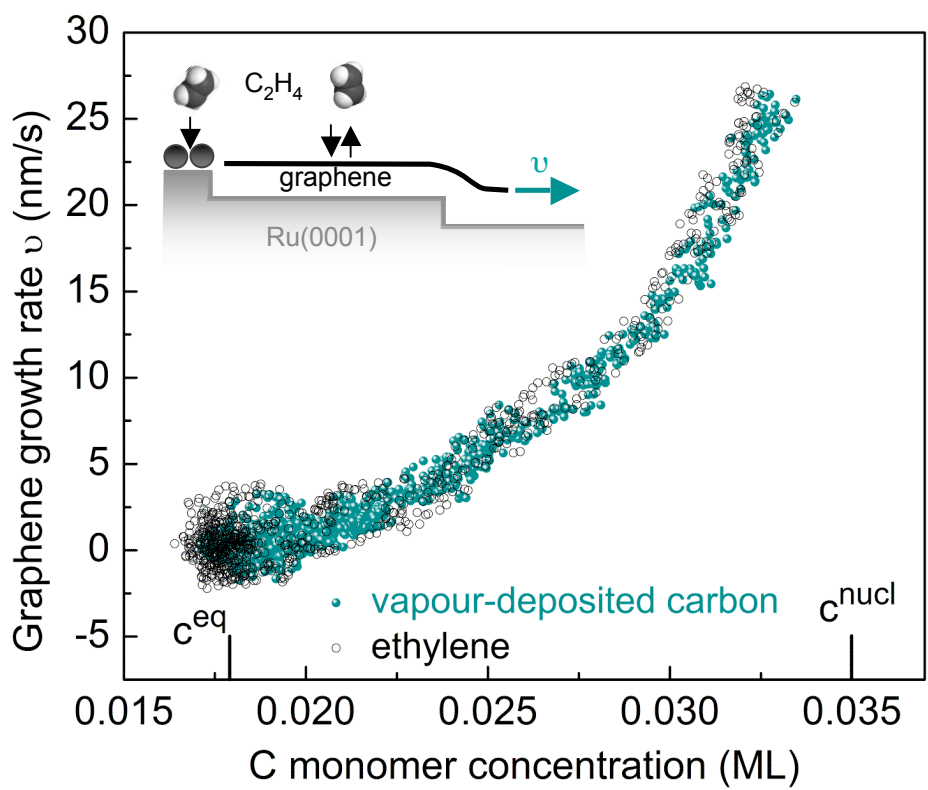

Figure 6. Growth rates of graphene islands $\upsilon$ as functions of C monomer concentrations for vapor-deposited carbon (green filled circles) and for ethylene deposition (black hollow circles) on Ru(0001) at 1020 K. Insert illustrates definition of the graphene growth rate $\upsilon$ as the linear velocity at which the graphene edge advances.

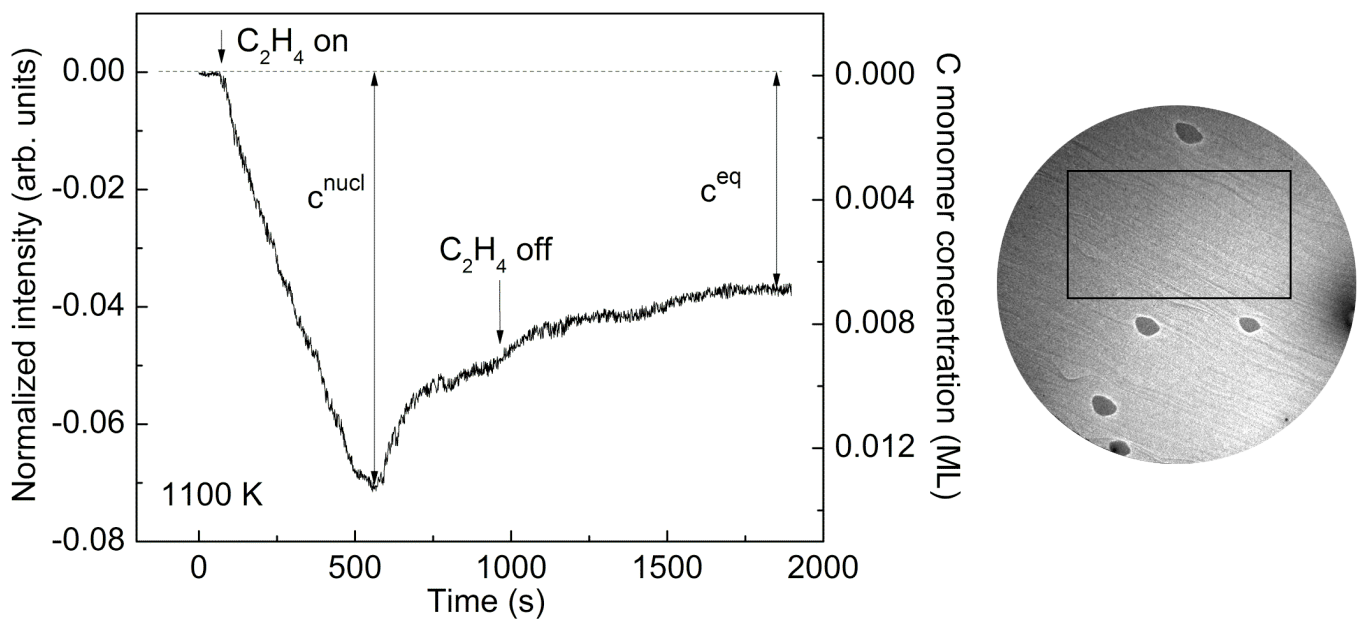

Figure 7. Electron reflectivity (left axis) from Ir(111) and calibrated carbon monomer concentration (right axis) dependence on ethylene exposure time at pressure 8×10$^{-10}$ Torr and temperature 1100 K. LEEM image (46 μm field-of-view) is shown on the right panel taken at 1800s. The black box marks the graphene-free region, from which the C monomer concentration was determined.

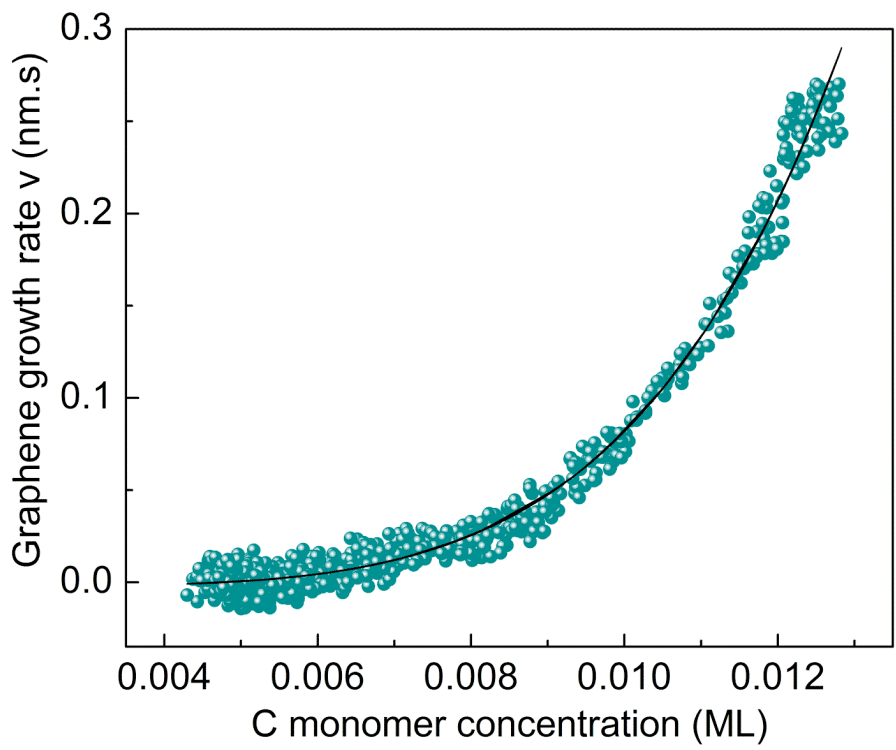

Figure 8. Graphene island growth rates dependence on C monomer concentration for ethylene deposition on Ir(111) at 1170 K. The solid black line is a fit to the equation of 5 C-atoms cluster attachment [26].

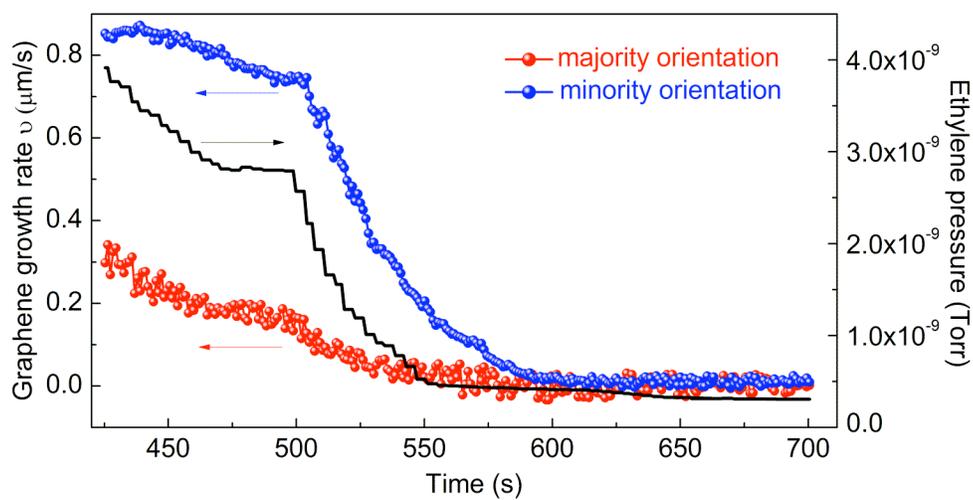

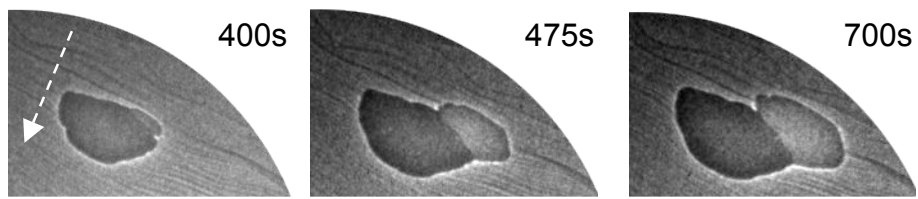

Figure 9. (a) Growth rates of two types of graphene islands on Ir(111) during ethylene deposition (left axis) with varied pressure (right axis) at 1150 K. Bottom panel shows LEEM images (24 μm × 18 μm) of two types of carbitic islands grown on Ir(111) at 400s, 475s, and 700s of $C_2H_4$ deposition. A white arrow indicates the "downhill" direction of Ir steps. The bright island nucleates at the perimeter of the dark island.

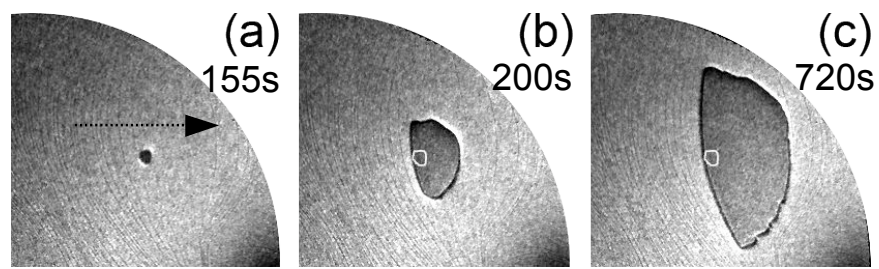

Figure 10. (a)-(c) Sequence of LEEM images (10 μm field-of-view) taken at 155s, 200s, and 720s during ethylene deposition on Ru(0001) at 960 K. A white contour in (b) and (c) indicates the position of the nucleated graphene island in (a). A black arrow in (a) indicates the "downhill" direction of Ru steps.

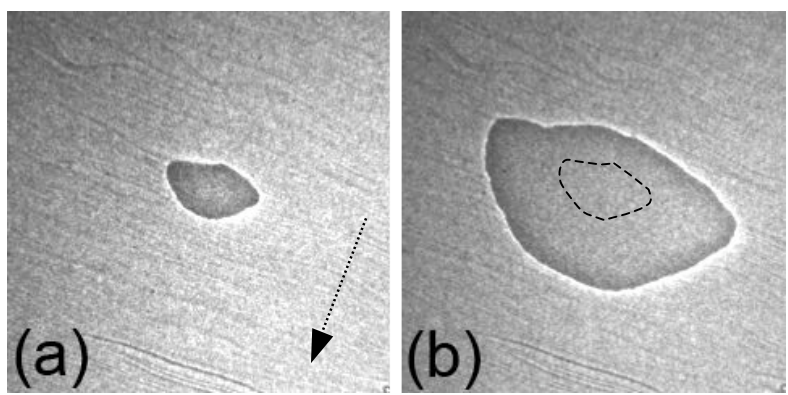

Figure 11. LEEM images (23 μm field-of-view) taken at the beginning (a) and at the end (b) of ethylene deposition on Ir(111) at 1200 K. A black arrow in (a) indicates the "downhill" direction of Ir steps. A black dashed contour in (b) indicates the position of the nucleated graphene island in (a).

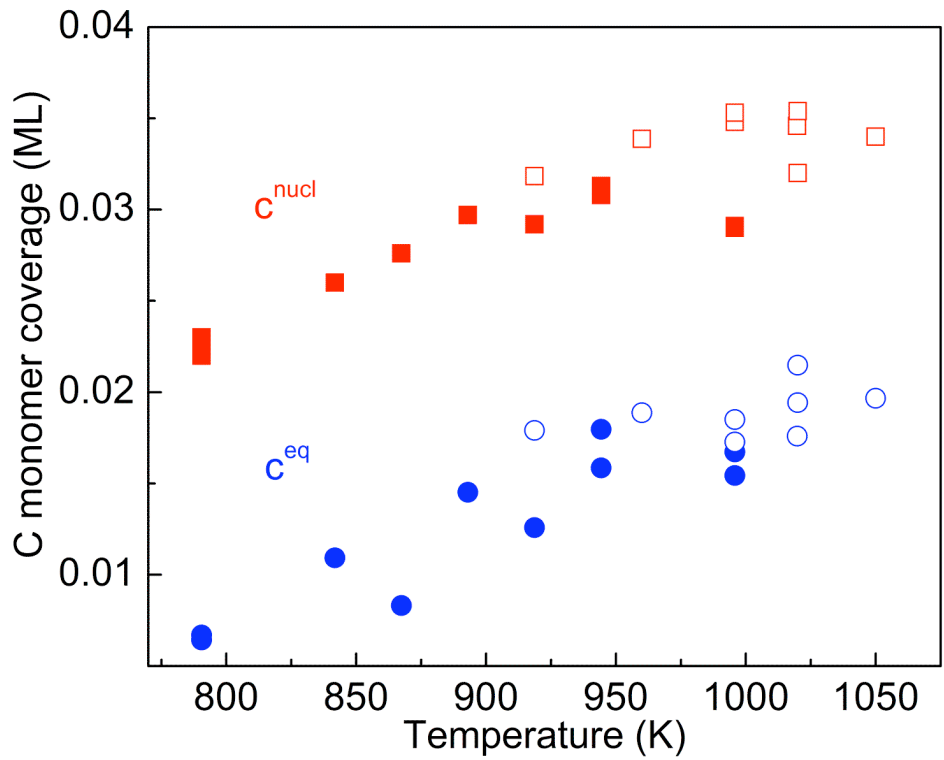

Figure 12. Temperature dependence of the C adatom concentration in equilibrium with graphene, $c^{eq}$ (circles), and C adatom concentration needed to nucleate graphene, $c^{nucl}$ (squares), on Ru(0001) after carbon-vapor (filled symbols) and ethylene (hollow symbols) depositions.